\documentclass[11pt]{article}
\pdfoutput=1

\usepackage{deauthor}
\usepackage{times}
\usepackage{wrapfig}
\usepackage{graphicx}
\usepackage{xspace}
\usepackage{cite}
\usepackage{amsmath,amssymb,amsfonts}
\usepackage{textcomp}
\usepackage{graphicx}
\usepackage{algorithm}
\usepackage{algorithmic}
\usepackage{xcolor}

\usepackage{txfonts}
\usepackage{url}

\usepackage{balance}  
\usepackage{booktabs} 
\usepackage{enumitem}
\usepackage{multirow}
\usepackage{subcaption}

\usepackage{listings}

\usepackage{outlines}
\usepackage{authblk}

\usepackage{etoolbox, totcount}
\newtoggle{shownotes}
\toggletrue{shownotes}
\newtotcounter{numnotes}
\newcommand{\note}[3]{
  \iftoggle{shownotes}{\stepcounter{numnotes}\textcolor{#1}{#2: #3}}{}
}

\newtoggle{showusernotes}
\newcommand{\joey}[1]{
    \iftoggle{showusernotes}{\note{cyan}{joey}{#1}}{}
}
\newcommand{\gabe}[1]{
    \iftoggle{showusernotes}{\note{magenta}{gabe}{#1}}{}
}

\title{CoVista: A Unified View on Privacy Sensitive Mobile Contact Tracing Effort}
\author{David Culler}
\author{Prabal Dutta}
\author{Gabe Fierro}
\author{Joseph E. Gonzalez}
\author{Nathan Pemberton}
\author{Johann Schleier-Smith}
\author{K. Shankari}
\author{Alvin Wan}
\author{Thomas Zachariah}
\affil{UC Berkeley}

\begin{document}

\maketitle

\begin{abstract}
Governments around the world have become increasingly frustrated with tech giants dictating public health policy. The software created by Apple and Google enables individuals to track their own potential exposure through collated exposure notifications. However, the same software prohibits location tracking, denying key information needed by public health officials for robust contract tracing. This information is needed to treat and isolate COVID-19 positive people, identify transmission hotspots, and protect against continued spread of infection.
In this article, we present two simple ideas: the \textbf{lighthouse} and the \textbf{covid-commons} that address the needs of public health authorities while preserving the privacy-sensitive goals of the Apple and google exposure notification protocols. 
\end{abstract}

\section{Introduction}

\begin{figure}[t]
    \centering
    \includegraphics[width=0.8\textwidth]{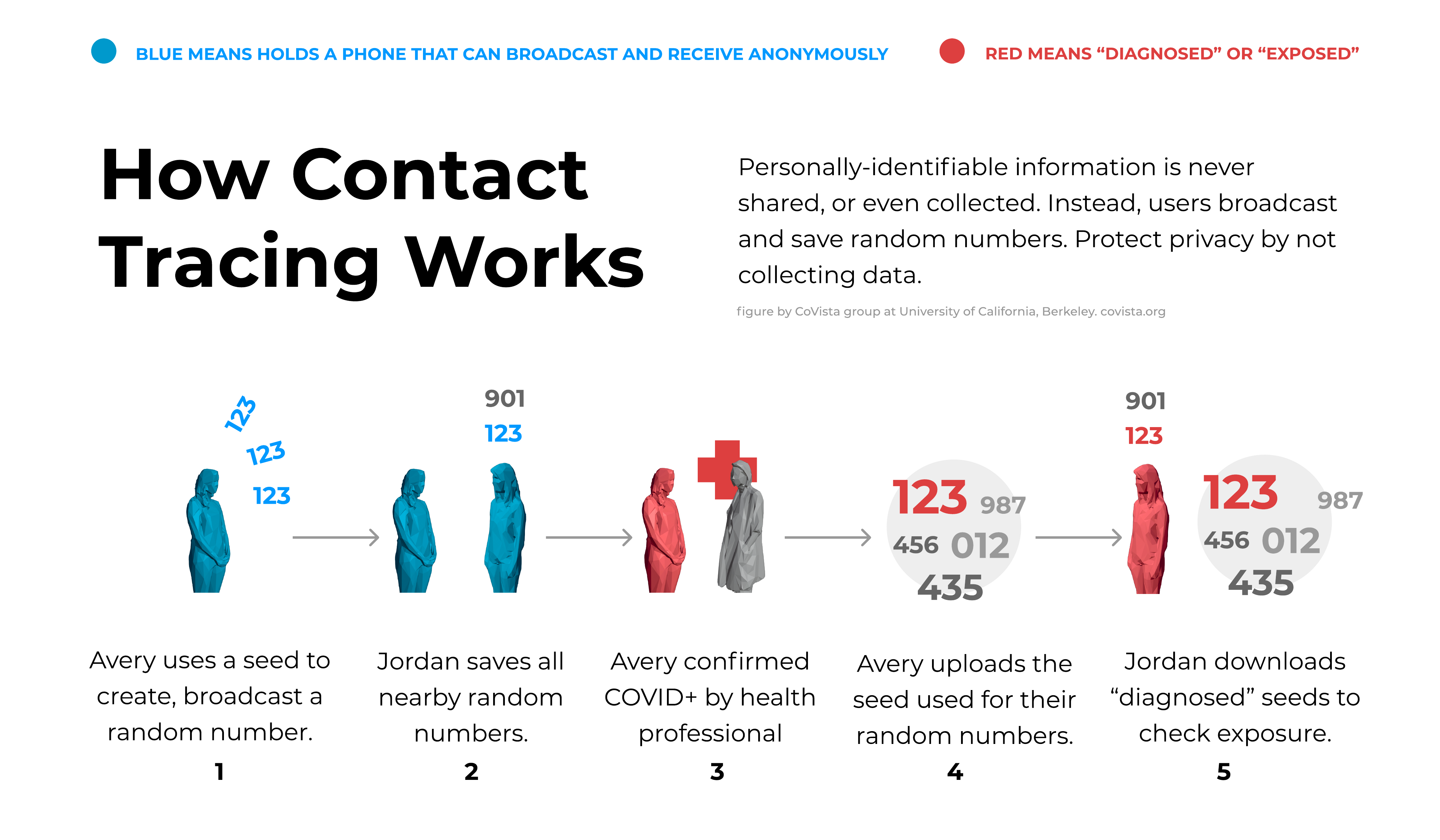}
    \caption{\textbf{Apple-Google Exposure Notification (AGEN) Protocol Overview.} }
    \label{fig:contact_tracing}
\end{figure}

Apple and Google have adopted a decentralized approach to mobile contact tracing that prioritizes individual privacy~\cite{agen}. Under the Apple-Google Exposure Notification (AGEN) protocol (see Fig.~\ref{fig:contact_tracing}), individual phones determine if the user has been exposed, without revealing \emph{the identity of the infected individual} and \emph{where the contact event took place}
The AGEN protocol is related to contemporaneously proposed protocols including PACT and DP-3t~\cite{pact,dp3t}.
Like these other protocols, the AGEN protocol does not use location information. 
Instead, it relies on the Bluetooth radios present on all modern phones to detect proximity with others. 
Beyond not collecting Protected Health Information (PHI), the decentralized approach retains the non-PHI on the phone, allowing individuals to determine risk locally on their device.

Governments and public health authorities want to understand where and how the disease is spreading, so they can take preventative measures.  
They also want to be able to use mobile contact tracing to augment existing manual contact tracing efforts.
With these goals in mind, governments advocate for a centralized approach, whether national or regional, where they maintain records of each person’s locations and interactions. 
This allows governments to determine exposures and notify people directly, as timeliness reduces spread. 
While centralized contact tracing may offer utility critical to re-opening the world’s economy, it raises profound concerns for civil liberties and personal privacy.
Government efforts that avoid reliance on the industrial Exposure Notification offerings have run into a host of failings, including reliability, power drain, interoperability, and participation.

Apple and Google have taken an unprecedented position -- essentially dictating public policy, not just by requiring the decentralized approach, but also by prohibiting contact tracing apps from collecting location information.
Further, they are restricting access to the new contact tracing APIs to national governments and permitting only one app per country or region. 
This decision circumvents the local governments, tribal organizations, and community health services that are often most aware of existing manual contact tracing efforts and the needs of their communities.  
Meanwhile, government’s contact tracing apps have failed due to restrictions imposed by AGEN.

In this article, we present two simple measures that enable the AGEN protocol to support manual contact tracing efforts, provide visibility into the spread of disease, and return authority to local communities all while preserving privacy within the Apple and Google framework. 
\begin{enumerate}
\item \textbf{Treat places as people.} Endow public places with the same privacy-preserving technology we used to monitor exposure for individuals.
\item  \textbf{Nation-scale data, not apps and processes.} Build a common backend for the AGEN protocol that spans apps and governmental boundaries.
\end{enumerate}

In the rest of this article, we describe these two simple measures and how they both improve contact tracing while also preserving individual privacy.

\subsection*{Lighthouse: Treat Places as People}
\begin{figure}[h]
    \centering
    \includegraphics[width=\textwidth]{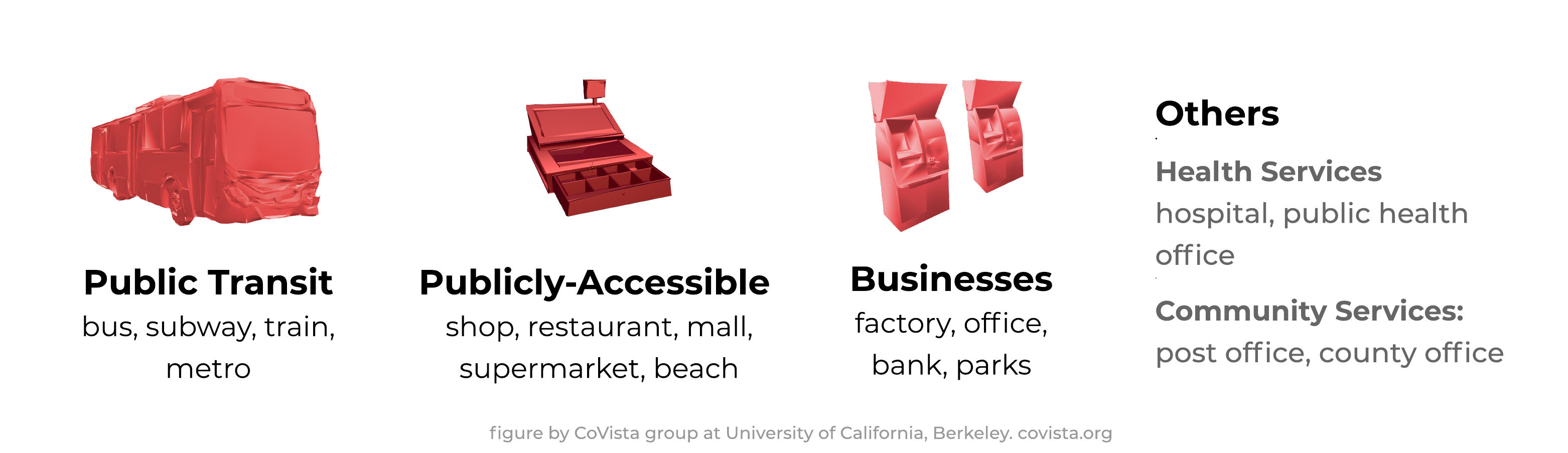}
    \caption{\textbf{Lighthouses Overview.} Lighthouses can extend the AGEN protocol to physical places}
    \label{fig:placesExamples}
\end{figure}

If we treat public places as people, we can use the AGEN protocol to (a) understand COVID-19 exposures across space, (b) integrate with manual contact tracing, and (c) do so with the same privacy-sensitive protocol. To treat places as people in AGEN, simply attach mobile phones or specialized low-cost beacons to publicly accessible places (e.g., county services, stores, buses).  
Like a lighthouse, these devices help communicate risk associated with places.  
Well-positioned, they can offer robust proximity detection, can detect their exposure, and the can convey aspects the risk that represents.

By choosing to share their locally computed exposure risk with public health authorities through the AGEN protocol, owners of publicly-accessible places can aid in mitigating virus spread. 
Alternatively, if a place is identified through traditional, manual contact tracing, the place can still anonymously participate in the AGEN protocol, notifying others without revealing where they were exposed. 
Treating places as people empowers stewards of public spaces to collaborate with public health authorities to help mitigate the spread of disease without jeopardizing the privacy of patrons or the reputation of the public spaces.
This procedure can facilitate detection of exposure from a non-participating individual while improving anonymity over manual contact tracing methods.  Going even further, such places could provide other means of beaconing that do not involve smartphones, such as QR code displays, codes on receipts and so on.

\subsection*{COVID Commons: A Nation-scale Data Backend}

Rather than “one app per nation,” a better solution would be to provide a common privacy-preserving data exchange across apps and administrative boundaries — a Commons.  This would allow societal structures and innovation, rather than corporate policy, to determine how the app ecosystem should evolve.  It is very likely that participation will be greatest if the apps are available through local organizations (e.g., tribal organization, university campus) that individuals trust.   A common privacy-preserving data exchange is already compatible with the AGEN protocol.

\begin{figure}[h]
    \centering
    \includegraphics[width=\textwidth]{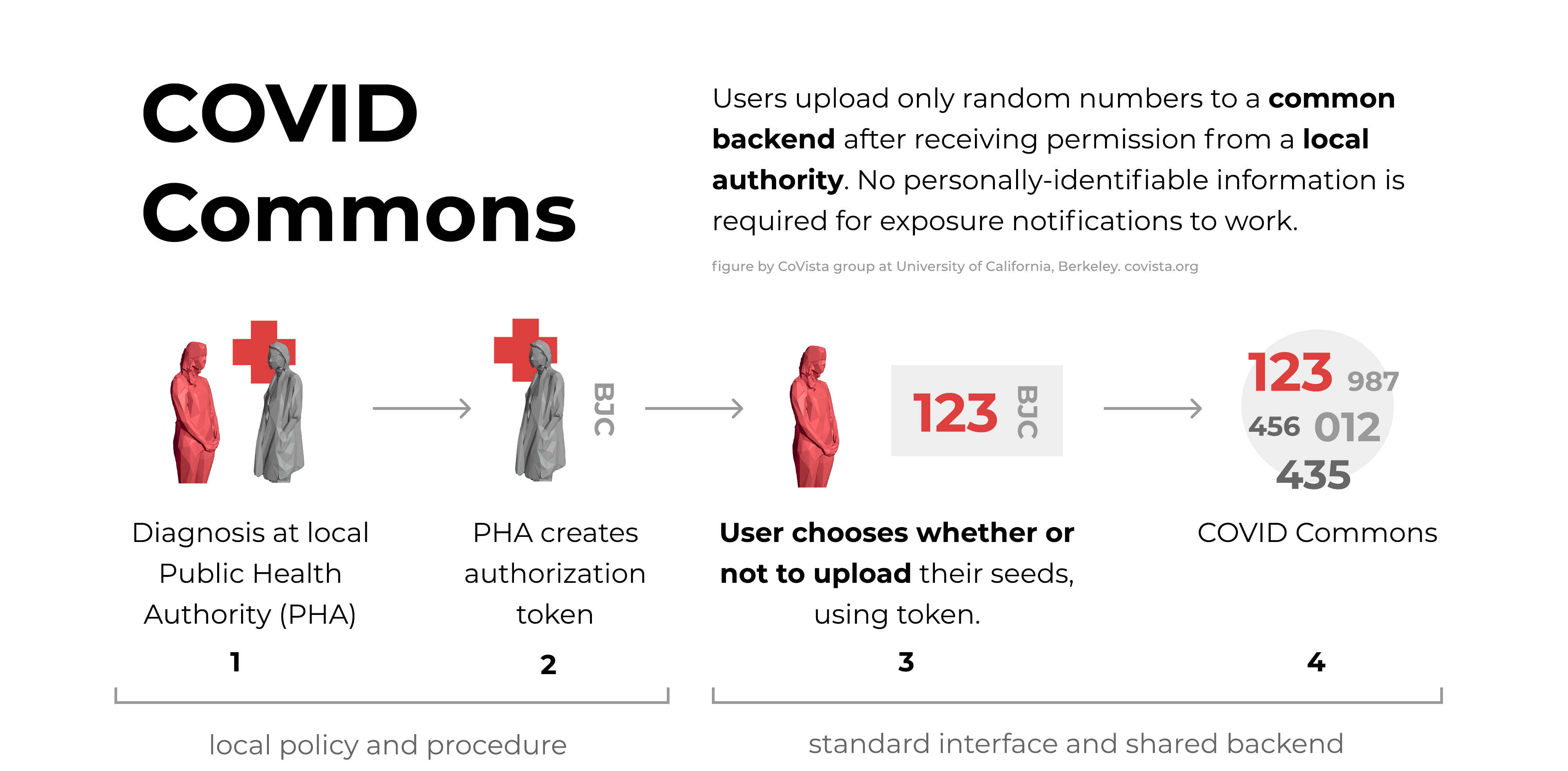}
    \caption{Caption}
    \label{fig:commoms}
\end{figure}

When an individual tests positive and they engage in a conventional contact tracing interview with a public health professional, the professional obtains an authorization so the individual, on an opt-in basis, can share their anonymous exposure information.  Public health professionals serve to protect the integrity of the information in the Commons without exposing any patient data or medical data. 

Their actions are quite similar to publishing counts of cases, statistics and demographic information, as is done today.  The Commons might be hosted by governmental or NGO structures, based on national or regional policy.  A diverse and innovative app ecosystem can grow to meet the needs of individuals and agencies.

In the remainder of this article we describe both these technical solutions in greater detail.  
We have organized each section to be relatively self-contained.

\section{Background on Privacy-Sensitive Mobile Contact Tracing}

The key building block for Privacy-Sensitive Mobile Contact Tracing (PS-MCT) is a subtle combination of radio protocols, cryptography, and risk-calculation.  
Phones have a short-range radio, Bluetooth, used to connect to nearby devices.  
To make those connections it periodically broadcasts tiny bits of information. 
The PS-MCT protocols leverages this short-range background broadcast to resolve nearby individuals.

In the Apple-Google Exposure Notification (AGEN) protocol, each phone generates a daily secret key called a Temporary Exposure Key (TEK).
Then every 15 minutes the phone uses the TEK to generate a new 16 byte Rolling Proximity Identifier (RPI). The RPI sequence is generated using a crytographic hash function and it therefore caries no information about the source individual.  
The current RPI is then continuously broadcast every few hundered milliseconds.
All phones log the RPIs they hear for future exposure analysis.  
Because the RPI is continuously changing, it also cannot be easily tracked.

When someone tests positive they can \textbf{anonymously} publish the daily keys (TEKs) from the days when they were contagious.  
The confirmed positive collection of TEKs is called a Diagnosis Key in the AGEN protocol.  
The Diagnois Keys are published by sharing them with a trusted and server whcich publishes the TEKs for download.
Others can obtain these keys and use the same cryptographic hash functions to recreate the sequence of RPIs and determine if they encountered any infected individuals.  
This entire process is accomplished within the Android and iOS operating systems with government sanctioned apps being responsible for authenticating infected individuals and, with user permission, publishing the keys. 

It's important to note the distinction between policy and mechanism. The AGEN protocol (and the extensions proposed in this paper) provide a mechanism to detect and notify users about exposure risk, but it is up to public health authorities to define what constitutes an exposure. This distinction is explored in more detail in section \ref{sec:commons}.

\section{Lighthouses: Treating Places as People}


Despite their promise, there remains significant concern around the efficacy of privacy-preserving contact tracing and exposure notification protocols.
For one thing, not everyone has access to a mobile phone, and many that do may be unwilling to participate (even Singapore only achieved 20\% adoption\cite{traceTogether20pct}). 
This is a serious concern as the probability of a successful detection grows quadratically with participation~\cite{pact}. 
Of course, manual contact tracing does not have this issue and remains the gold standard. 
Will these two techniques exist in isolation or will they interact synergistically? 
Finally, bluetooth contact detection is limited in range and time; it cannot detect many important forms of transmission like surfaces or HVAC systems\cite{hvacTransmission}.

\subsection{Treating Places as People}
There is a simple extension to mobile contact tracing that can help address these issues. 
The idea is rooted in centuries old maritime signaling. To navigate at night, ships use signal lights, much like our Bluetooth beacons, to identify and safely navigate around nearby ships. 
However, to safely sail at night, ships also rely on lighthouses, strategically placed beacons, to identify and safely navigate around key landmarks. 
It is this second form of beacon, the lighthouse, that is needed to address several of the key limitations in privacy-sensitive mobile contact tracing.

\subsection{What is an Exposure Notification Lighthouse?}
\begin{figure}[h]
    \centering
    \begin{subfigure}[t]{0.3\textwidth}
        \centering
        \includegraphics[width=0.9\textwidth]{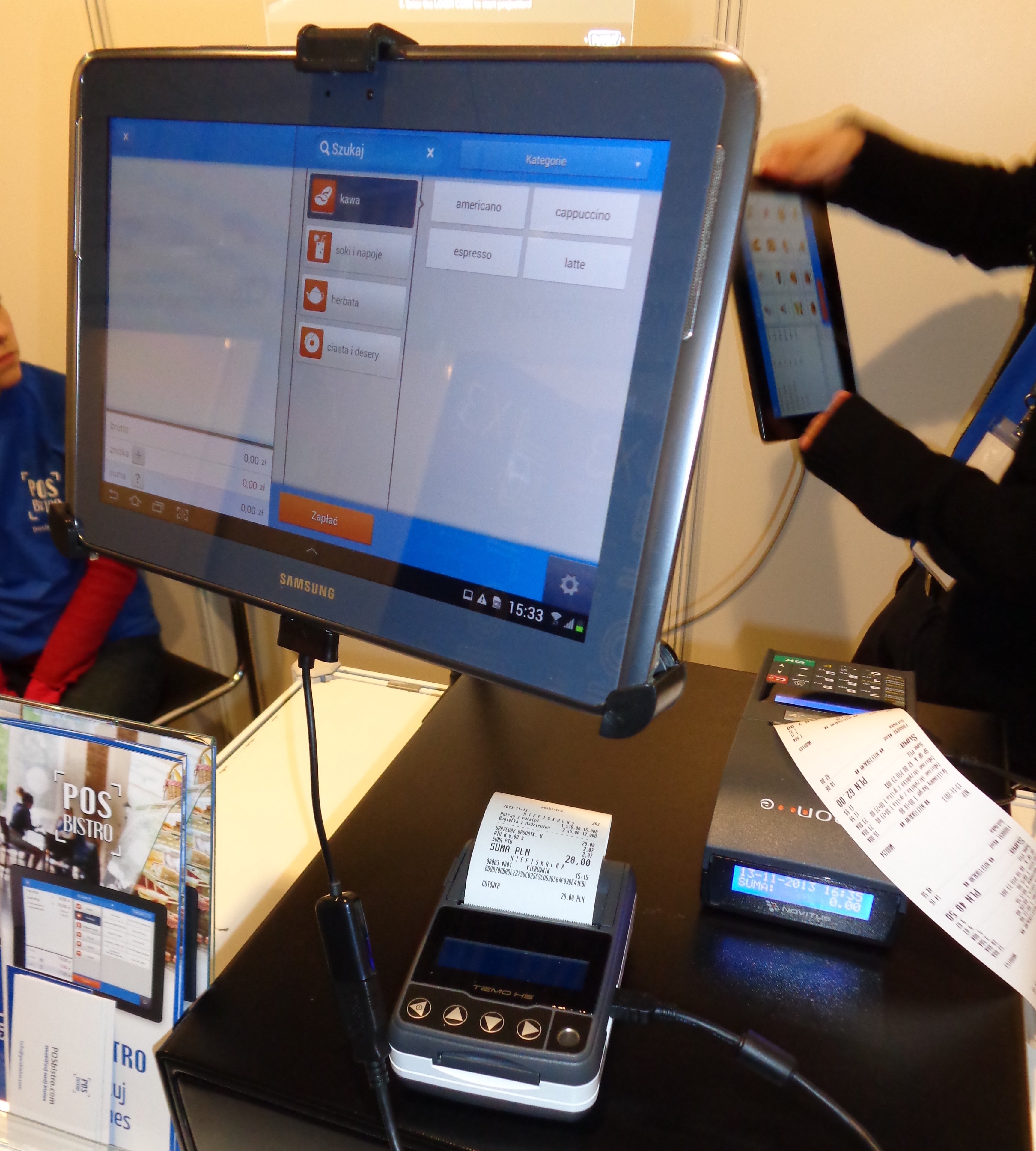}
        \caption{\textbf{Existing Devices:} Easily deployed immediately but expensive to deploy to new places. They can also be unreliable when unattended.}
        \footnotetext{wikimedia commons © Travelarz CC-ASA 3.0}
    \end{subfigure}
    ~
    \begin{subfigure}[t]{0.3\textwidth}
        \centering
        \includegraphics[width=0.9\textwidth]{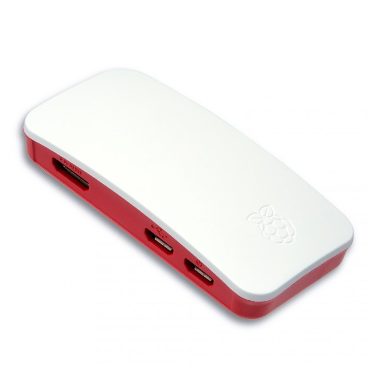}
        \caption{\textbf{Dedicated Devices:} Cheap and reliable. They will require new development and can be tricky to place correctly.}
    \end{subfigure}
    ~
    \begin{subfigure}[t]{0.3\textwidth}
        \centering
        \includegraphics[width=0.9\textwidth]{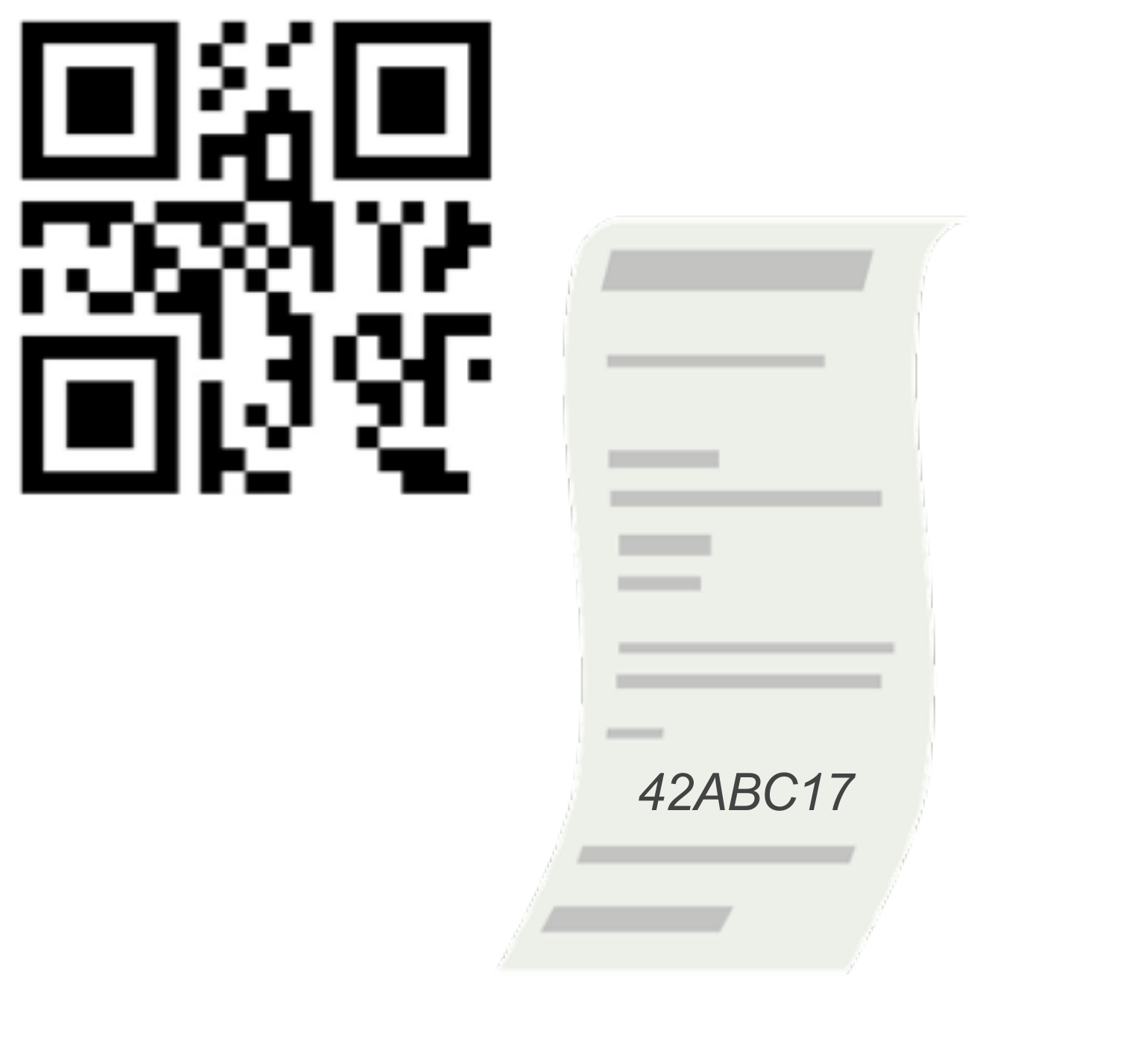}
        \caption{\textbf{No Device:} Allows those without the app or even a mobile device to participate. They may be easier to deidentify and require manual effort from users to check.}
    \end{subfigure}
    \caption{Lighthouses can be deployed using a number of different devices with varying tradeoffs. Each of these examples is feasible to deploy in the near future.}
    \label{fig:lighthouseDevices}
\end{figure}

A privacy-sensitive exposure notification lighthouse is a device (e.g., a mobile phone or even a smart sticker) deployed in a public space following the same privacy sensitive mechanisms as individuals. Figure \ref{fig:lighthouseDevices} gives a few examples of what lighthouses could look like.
Much like the maritime lighthouse, contact tracing lighthouses can be used to inform others of potential exposures associated with public spaces discovered through manual contact tracing. 
A contact tracing lighthouse can also log passing beacons to inform owners and public health authorities of exposure risks. However, because the lighthouse follows the same privacy-sensitive protocols as individuals, it retains all of the privacy guarantees of the existing protocols. Moreover, by installing contact tracing lighthouses in participating public spaces ranging from stores and restaurants to schools and buses, we introduce a privacy sensitive mechanism to bridge manual contact tracing with mobile contact tracing while also giving public health authorities the ability to gain visibility into the spread of disease while preserving individual privacy.

\subsection{What Do We Get From Lighthouses?}
\begin{figure}
    \centering
    \includegraphics[width=\textwidth]{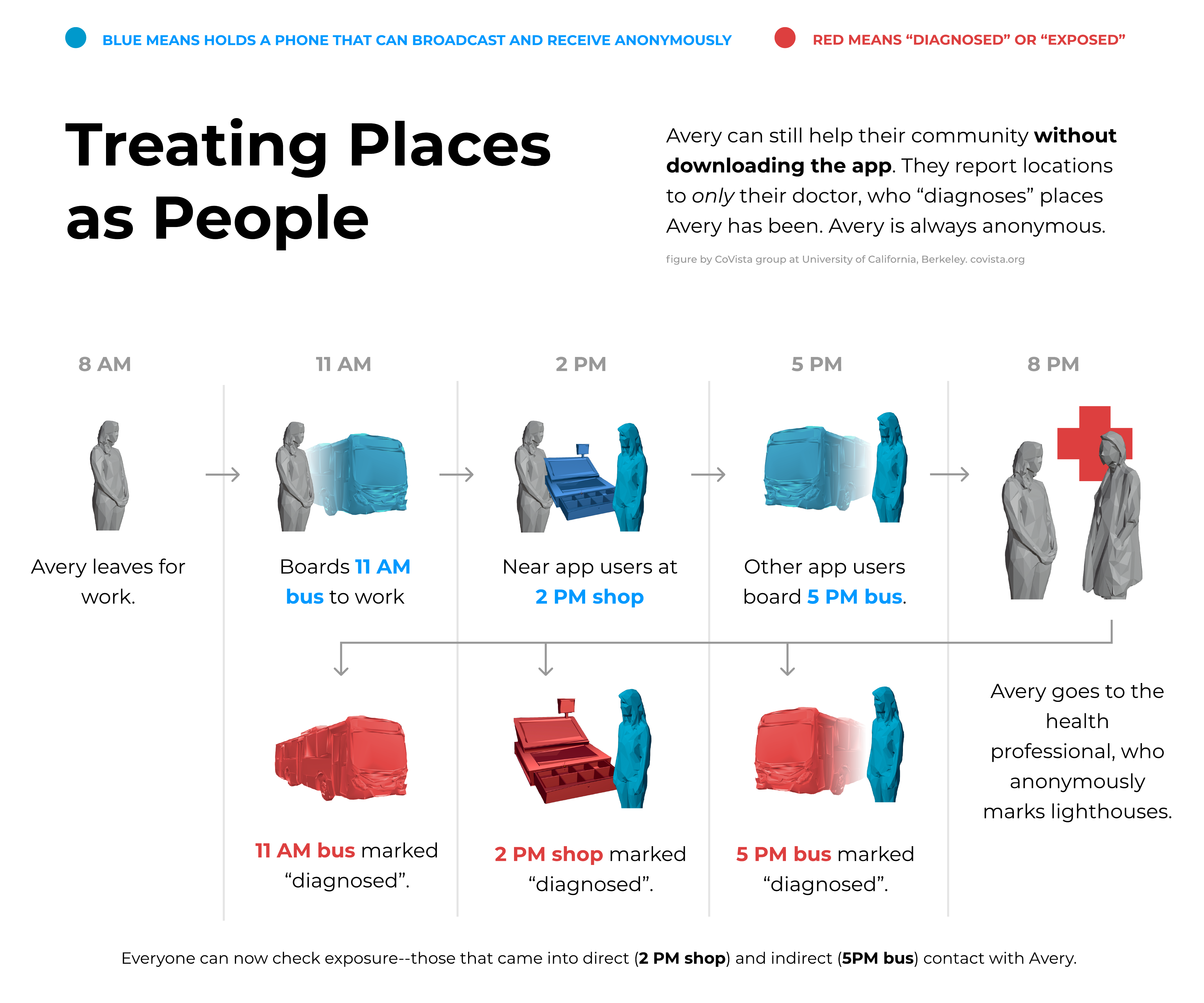}
    \caption{An example scenario of an affected individual (Avery) and a stranger (Bernie) that may have been exposed. Even if Avery is not an app user, the doctor can still notify the shop and bus of their exposure; lighthouses \textbf{connect manual and mobile contact tracing}. Lighthouses \textbf{bridge place and mobile} by allowing the bus and shop to notice their exposure and mark themselves as exposed, even if Avery did not remember visiting them. They also \textbf{enabled less direct forms of exposure} by notifying Bernie of there potential exposure, even though they were not on the bus at the same time as Avery. Finally, lighthouses \textbf{went beyond the standard mobile contact tracing} benefits because the shop and bus can track how often they are exposed, at what times, and even where in their location (if they have multiple lighthouses installed).}
    \label{fig:lighthouseExample}
\end{figure}

Figure \ref{fig:lighthouseExample} walks through an example of how lighthouses could work in a typical case. Let's explore some of these benefits in more detail.

\subsubsection{Bridging Place and Mobile}
The contact tracing lighthouse empowers stewards of public spaces (e.g., shop owners, school administrators) to collaborate with public health authorities to help mitigate the spread of disease without jeopardizing the privacy of patrons or the reputation of the public spaces. When an individual is tested and confirmed, the manual contact tracing interview process begins. One of the first questions asked is “Can you recall where you have been that might have exposed others?”. A place, unlike an individual, has a large set of explicit relationships with its institutional environment - business license, health department approvals, chamber of commerce, etc. The interview process routinely seeks to gather information about places. But often there is little the place can do to help. With its lighthouse, it has a very simple way to provide assistance without undue impact on its reputation. Just like a person, it can check its own potential exposures. And, like a person, it can anonymously publish its own random numbers as COVID positive so that people can use them for detecting their exposure risk. This is especially useful if the individual who tested positive was not using an app that broadcasts their own sequence of random numbers. Unlike a person’s phone, stationary lighthouses can be carefully positioned to avoid issues like weak or unreliable broadcast that mobile phones can face. Finally, the public place may also take other measures in assisting its local health authority in detecting hot spots, such as informing them of the exposure risk that it observes.

\subsubsection{Indirect Transmission}
The lighthouse can be used to address forms of indirect transmission that are not captured in the existing privacy-sensitive mobile contact tracing efforts. If a COVID-positive individual enters a bus and touches or sneezes on several surfaces they could potentially infect others over the next several hours, long after they have left the bus. Air conditioning systems have also been implicated in spreading the virus over large distances~\cite{hvacTransmission}. Existing wireless protocols will fail to capture these forms of transmission since they rely on close, contemporaneous radio proximity with the positive individual. However, with the introduction of the lighthouse, the bus or restaurant can automatically determine that it was exposed and choose to anonymously publish its own random number sequence.

\subsubsection{Beyond Mobile Contact Tracing}
Finally, the lighthouse can also extend the privacy-sensitive mobile contact tracing protocol beyond the smartphone. Places provide channels of communication to individuals that complement smartphones. Simple lighthouse codes might be included on printed receipts or handed to customers so that a person without contact tracing on their phone might use such code they have received to privately check for exposure on public websites somewhat like they would if they used such an app.

\subsection{Lighthouse Privacy}
There are legitimate privacy interests for places, as there are for people. 
Businesses may fear irrational boycotts or loss of reputation, and there are risks of perpetuating prejudice or stigma against neighborhoods or ethnic groups (already a growing problem \cite{asianDiscrimination}). The good news is that lighthouses inherit the privacy protecting characteristics of the AGEN protocol. In their simplest form, they are phones running ordinary apps, different only because they live in a fixed place rather than in purse or pocket.

Lighthouses can complement manual contact tracing by offering greater confidentiality in situations where people cross paths in stigmatized locations. For example, investigators following up on a recent case cluster associated with nightclubs in Korea catering to the LGBTQ community have struggled to contact attendees who are afraid of being “outed” or facing discrimination \cite{koreaClub}. Lighthouses can fill such gaps without requiring or relying on anyone, even an affected person, to disclose their location history.

A potential concern about lighthouses is that internal state, if combined and aggregated, could be used to undermine some of the privacy guarantees of AGEN. 
Indeed, with a sufficiently large deployment of lighthouses and centralized collection of recieved RPIs, it would be possible to resolve the locations of all the COVID positive individuals.
We therefore stress that each lighthouse must follow the existing decentralized protocol for risk assessment in which only risk is aggregated and not the underlying RPIs.  

To address some of these security concerns, it is possible to deploy \textbf{transmit only} (passive) lighthouses which could be used to anonymously notify others of exposure without listening for RPIs.  
This eliminates the risk of resolving COVID positive individuals locations but also limits visibility into where the disease is spreading.

\section{Covid Commons: Unified Contact Tracing With Many Apps}
\label{sec:commons}


Standardizing on an exposure notification protocol such as AGEN is crucial to achieving the critical mass of participation required to make such an approach effective (estimated to be at least 80\% for a city of 1 million individuals~\cite{hinch2020effective}) and privacy preserving.
Apple and Google, as producers of the two dominant smartphone OSes, are uniquely positioned to bring these capabilities to nearly every smartphone and have engineered the AGEN protocol to that end.
Providing an implementation of the protocol through an OS upgrade would eliminate the fragmentation of the protocol preventing sufficient adoption.
However, standardizing the proximity detection protocol is just part of the solution.

Due to concerns over potential misuse of the capabilities of the proximity detection protocol and location tracking features of modern smartphones, Apple and Google have announced that access to the SDK for AGEN to ``public health agencies"~\cite{agen_pha_announce}.
The realization of this policy has evolved to essentially mean ``one app per state."
Centralizing the apps around state governments means that tracing efforts are not limited to local jurisdictional boundaries. 
However, the public health authorities responsible for testing and tracing often operate at a more local level~\cite{bay_area_testing, local_testing_sites}.


The fundamental issue with this policy is a failure to distinguish between ownership of the required repository of ``diagnosis keys'' and the applications that produce and utilize them.
The repository contains keys that correspond to times when diagnosed individuals may have interacted with others and must therefore be public-facing so that individuals, via AGEN-compatible apps, can download the keys and determine their own exposure risk.
The keys uploaded to the repository must be limited to those that have been authorized by a public health authority through some testing and interview process.
In the U.S., public health measures such as contact tracing and case reporting are carried out at a county or city level through individually executed processes, in coordination with state and national law and policy.
However, health authorities do not generally have the technical, human or financial resources to produce the required app and also stand up and maintain such a public-facing data service.
This is even less tenable during a pandemic, when such resources are already strained.
Who, then, manages the AGEN backend needed to share Report Keys?

The worst-case scenario is for the repositories to emerge as fragmented silos of diagnosis keys.
In the midst of the custody battle over administration of the emerging technical approaches to the contact tracing problem, it is important to remember that exposure does not respect administrative boundaries.
Regardless of how the repository is established, it is essential that exposure notification can be transmitted across backends and across apps.
Indeed, there are already efforts to combine contact tracing efforts and data across the administrative boundaries established by industry's policy around AGEN~\cite{tri_state_coalition}.

\subsection{A Path to Nation-scale Data}

A solution to this issue is the creation of a privacy-preserving data exchange shared and accessible across apps and administrative boundaries --- a Commons.
Different public health authorities could cooperatively contribute to the Commons and individuals across many jurisdictions would access it through their apps.
Such a Commons might operate at the scale of one per nation, or it might be regional with some form of federation to share keys and exposure information across individual instances.
The Commons may be hosted and maintained by governmental authorities, foundations or other appropriate institutional entities.

The Commons requires no change to the EN protocols.
Each individual participates in the proximity detection using daily TEKs and the RPIs generated from them without change.
With no other information leaving the phone, the app on an indivual phone downloads the diagnosis keys and presents them to the SDK to obtain exposure risk scores.
The difference is that those diagnosis keys are not \emph{a priori} limited to the ones produced by users of the same app.
Just as in the current decentralized protocol, an individual who tests positive --- i.e. a confirmed case --- and participates in contact tracing with a specific public health authority is asked to voluntarily submit their TEKs for the past period over which they are likely to have been contagious.
The AGEN protocol does not stipulate how that request is formulated or how that submission is performed, but the introduction of the Commons allows us to answer that question precisely.

\subsection{The PHA Experience: Federating Access to the Commons}

A Commons is utilized by a well-defined set of public health authorities that are registered with it.
Professionals at PHAs can authenticate to and access the Commons using well-established authentication techniques such as LDAP and OAuth.
As part of requesting a patient to submit their diagnosis keys to the Commons, the professional requests a one-time authorization (OTA) from the Commons.
The OTA authorizes the upload of TEKs corresponding to the extent of time during which the patient is likely to have been contagious; this will involve some days prior to the diagnosis and extend to several days or weeks following.
The OTA is provided to the patient to authorize the submission of their keys to the Commons; the upload is opt-in, as under the usual AGEN protocol.

Because the OTA corresponds to a single interaction between a healthcare professional and a patient, it is a natural key on which helpful metadata can be associated.
The authorizing PHA may want their identity to be associated with the submitted diagnosis keys so that the provenance of the diagnosis is preserved.
This identity, realized by the TLS public key of the PHA, can be associated with generated OTAs in the Commons.
This provides no more information than having each PHA host their own exposure key store, as currently envisioned.
As we will discuss below, the PHA identity can also be used to filter which diagnosis keys are considered in the matching process on an individual's phone.

The OTA itself is also a useful piece of metadata when associated with the uploaded keys in the Commons.
The authorizing professional will likely want to be able to determine whether the patient has in fact contributed their keys as requested to determine if follow-up is necessary.
Associating the OTA with the uploaded keys places no new information in the Commons: the professional knows what they know about those they interview and they know what authorizations they have requested.
Each such confirmed case can be expected to have generated some authorization request, but that information, along with all other aspects of the contact tracing process, is under the purview of the PHA.
Counts of confirmed cases and other aggregate information are already regularly reported to the public.

The Commons is able to provide this functionality despite no patient data --- indeed, no health or medical data of any kind --- ever being entered into the Commons.
The Commons also takes no position on which parts of the contact tracing process are automated and which are manual.
It integrates cleanly with the existing processes established by PHAs for interviewing patients, deciding which diagnosis keys are appropiate to share, and so on.
The Commons merely provides a means of carrying out the result of these decisions in a way that does not require individuals to be using the same phone app.

\subsection{The User Experience: Federating Downloads from the Commons}

\gabe{How much do we want to say here?}

The current draft of the AGEN protocol says very little about how exposure notification information will be transmitted or relayed among the different instances of the diagnosis key repository.
A unified Commons immediately addresses this issue --- all diagnosis keys are uploaded and downloaded from the shared repository.
However, under a federated regime, it is necessary to address how exposure notification and uploaded keys can be routed across instances of the Commons.

Consider a scenario where an individual is diagnosed by one PHA but may travel or may have traveled to regions covered by different PHAs.
Individuals in those other regions should be able to obtain the TEKs for the diagnosed individual, even though the diagnosis was performed by a remote PHA.
There are two complementary approaches.

The first approach proactively forwards diagnosis keys to relevant PHAs.
A PHA professional can ``tag'' a requested OTA with public keys or other metadata identifying relevant remote PHAs, using information acquired as part of the interaction with the individual.
The process of uploading diagnosis keys authorized by that OTA to the Commons can then incorporate a simple forwarding mechanism that replicates those keys to the federated instances of the Commons managed by those other PHAs.
This can be performed by the Commons itself, or it may be performed on an opt-in basis by the individual.

Under the second approach, individuals may proactively request diagnosis key downloads from federated instances of the Commons they are interested in.
An individual may subscribe to diagnosis key downloads for Commons covering regions the individual has travelled to or will travel to.
Additionally, an individual may subscribe to Commons for surrounding regions.

Regardless of whether the Commons is realized as an administratively centralized server or a federated mesh, the Commons must provide some mechanism for filtering or scoping the download of diagnosis keys onto an individual's device.
The number of keys downloaded from the Commons will grow with the adoption of AGEN-based apps, improvements in testing, and the spread of COVID-19.
Without the ability to filter the set of keys down to a reasonable and relevant set that is still large enough to maintain the privacy-preserving properties of AGEN, the bandwidth requirements and compute requirements (for deriving the thousands of RPIs used for the matching process) may grow to be untenable.
\joey{Can we do a small back of the envelope calculation here.}

\subsection{Extending the Commons}

Once this basic separation of key repository has been established, along with the method for authorizing submissions and routing them to queries, it becomes possible to meaningfully entertain the question of including earlier stages. 
Especially with testing being scarce, confirmation occurs late and involves the contact tracing processes of public health authorities.
Prior to that stage, we have Probable Covid (where a diagnosis has been performed based on defined symptom criteria), preceded by Suspected Covid, and often preceded by individual state of concern. 
Associated with each stage is a process, an (increasingly large) set of principals who might authorize a submission, and a reduced significance in conveying exposure.
For example, the physician or health service making a Probable Covid diagnosis (and typically also authorizing testing) would be the natural principal to authorize the individual to submit “Probable Keys”. 
The privacy-sensitive protocol could access these, as well as the “Confirmed Keys” referred to as Diagnosis Keys in the EN protocol. 
The protocol permits the distinction to be reflected in metadata carried along with the keys. 
Thus they might factor in, perhaps with lesser weight, into the risk score.

\section{Conclusion}
With these two simple innovations, we can move past the conflicts and controversy and on to utilizing privacy-sensitive mobile contact tracing to improve the efficacy of public health measures - thereby saving lives and unnecessary suffering - while respecting civil liberty.  The companies should provide the interoperable building blocks without themselves getting into the business of providing the Apps or holding the data.  They can maintain their privacy-first, decentralization posture, but should advocate for an interoperable covid key Commons, rather than dictating policy on the App ecosystem.  Government actors can regain policy determination and influence the App ecosystem so as to best tailor offerings to their constituents, resolve how best to provide a Commons of appropriate scale, and utilize physical measures not unlike what they do with security cameras, traffic signals, inspections and other infrastructure to appropriately related anonymous key reports to the places within their jurisdiction, recognizing the importance of public awareness, potential for creating stigma, and bringing benefit to non-participating members of the communities.  The technology can assist the contact tracing process, but should not be dictating it or replacing the human relationship of the patient and the health professionals performing interviews and care.  The trust that is built there is what makes opt-in approaches viable, and only with appropriate individual protections.

\bibliographystyle{IEEEtran}
\bibliography{ms}

\end{document}